\documentclass{aastex6}

\begin{document}

\title{The Imprecise Search for Habitability}

\author{Kevin Heng\altaffilmark{1}}
\altaffiltext{1}{University of Bern, Center for Space and Habitability, Sidlerstrasse 5, CH-3012, Bern, Switzerland.  Email: kevin.heng@csh.unibe.ch}

\begin{abstract}
Hunting for alien habitats without defining life? \\
\scriptsize
Kevin Heng is a professor of astronomy and planetary science at the University of Bern, Switzerland, the director of its Center for Space and Habitability, a subproject leader of the Swiss PlanetS National Center of Competence in Research, and a core science team member of the Swiss-led CHEOPS mission of ESA to hunt for Earth-like exoplanets around nearby stars.\\
\textit{(Published in American Scientist, May/June 2016 Issue, Volume 104, Pages 146--149.)}
\end{abstract}


\section{Introduction}

As planets are being discovered around other stars by the thousands, several scientific disciplines that traditionally exist in parallel are converging, including astronomy, planetary science, and biochemistry. They share an ultimate goal: to locate and identify life elsewhere in the Universe. We are engaged in a search for habitability---conditions suitable for life---despite missing a clear definition of what life is. We are hunting for what we cannot yet sharply define. Nevertheless, we can make some informed inferences about what life needs. From our one example of life discovered so far, it appears that liquid water is an essential ingredient. If an exoplanet orbits at the appropriate range of distances from its star to allow liquid water to exist on its surface, then it is said to be in the habitable zone---not too hot, not too cold, purportedly just right for living things.

The notion of the habitable zone has its origins in the planetary trio of Venus, Earth, and Mars in our Solar System. With surface temperatures in excess of 400 degrees Celsius, present-day Venus is a scorching inferno largely devoid of water. Its hostile temperatures are a direct consequence of its extremely thick atmosphere being dominated by carbon dioxide---a powerful greenhouse gas that makes up less than one part in a thousand of the atmosphere of Earth. Mars is a study in contrasts, having a thin atmosphere, large temperature swings, and an average surface temperature that is well below the freezing point of water. Earth sits conveniently between these two extremes. It is convenient to visualize Earth as residing within a zone of habitability, flanked by Venus and Mars. 

With the flurry of recent discoveries by the Kepler Space Telescope, it is routine to encounter media reports of ``habitable-zone exoplanets"---sometimes accompanied by speculation on what types of life forms may exist on them---using a conception of the habitable zone that extrapolates directly from that of our own Solar System. The habitable zone is a star-specific concept, however. Stars exist in a variety of sizes and masses. More massive stars tend to burn more brightly, but have shorter lifetimes. The most common types of stars in the Universe are not like our Sun, but have masses between 10 and 50 percent of it. These \textit{red dwarfs} have cooler temperatures than our Sun and radiate far less energy, which means that if a planet of Earth's size were to maintain the same range of atmospheric temperatures it would have to orbit such stars more closely. 

\section{A Problem of the Atmosphere}

It is somewhat harder to imagine that the habitable zone is also an atmosphere-specific concept. At the very minimum, it requires the specification of three types of atmospheric gases that strongly influence a body's surface temperature. First, we need an \textit{incondensible greenhouse gas}---one that stays in its gaseous form over the range of temperatures found in the atmosphere. On Earth, this role is played by carbon dioxide. Second, we need a \textit{condensible greenhouse gas}, which is able to exist in both its gaseous and liquid forms. Water is the condensible greenhouse gas of our atmosphere and the lynchpin of the hydrological cycle. 

The boundaries of the habitable zone are determined by what happens to the condensible and incondensible greenhouse gases at different distances from the parent star. The inner boundary occurs where the condensible greenhouse gas becomes incondensible, while the outer boundary occurs where the incondensible greenhouse gas becomes condensible. If the Earth were located too close to the Sun, then higher temperatures would result in more water existing as vapor, which in turn would lead to further warming. The planet would compensate for this greenhouse warming by emitting more infrared radiation and by shedding heat, but at some point there would be so much water vapor in the atmosphere that it would become opaque to infrared radiation\footnote{Known as the \textit{Komabayashi-Ingersoll limit}.}. At that point, the cooling of the atmosphere would be overwhelmed by heating, leading to a runaway greenhouse effect. Venus is believed to have suffered this fate during its history.

A third ingredient is an \textit{inert gas}, and its role is subtle. On Earth, the primary inert gas is molecular nitrogen. It does not contribute to greenhouse warming, because molecular nitrogen has an even distribution of electric charge across it\footnote{It is a symmetric diatomic molecule with no electric dipole moment.}. Quantum physics tells us that such molecules are largely incapable of absorbing radiation. Counter-intuitively, despite being the dominant gas by mass, molecular nitrogen is transparent to the radiation received and emitted by Earth. However, as the atmosphere heats up and accumulates more water vapor, water molecules participate in collisions with nitrogen molecules. Absorption of units of light or radiation, known as photons, must match the discrete energy levels within a water molecule. When water and nitrogen molecules collide, deficits or surpluses of energy are exchanged.  This phenomenon is known as \textit{pressure broadening}, and its overall effect is to increase the extent to which the water molecules may absorb radiation. Molecular nitrogen does not directly absorb light, but it influences how the greenhouse gases in the atmospheres do and thus  indirectly produces warming.

Inert gases also set a characteristic distance in the atmosphere known as the \textit{pressure scale height}, which determines if an atmosphere is puffy or compact. Hydrogen-dominated atmospheres tend to be puffier than their nitrogen-dominated counterparts. Furthermore, inert gases may participate in the chemistry involving greenhouse gases and alter their abundances.

If we were to move Earth farther away from the Sun, then at some point carbon dioxide starts to condense out of its atmosphere. As this greenhouse gas is removed, the atmosphere starts to cool and the overall temperature drops. The distance from the Sun at which the atmosphere becomes too cool to support liquid water on the surface of the body is the outer boundary of the habitable zone. At high pressures, molecular nitrogen may form transient pairs, which have an uneven distribution of electric charge across them\footnote{These pairs produce an electric quadrupole moment.}, that produce a weak greenhouse effect known as \textit{collision-induced absorption}. One imagines that the loss of gaseous carbon dioxide may be compensated by packing more molecular nitrogen into the atmosphere, but there is a limit to the mileage gained as it also condenses out, at some point, when the temperature becomes too low.

Once we understand how the condensible and incondensible greenhouse gases control the habitable-zone boundaries, we may imagine different flavors of habitable zones. Molecular nitrogen may be swapped out for molecular hydrogen, which has a considerably lower condensation temperature: tens of degrees Kelvin, rather than around a hundred. For planets with hydrogen-rich atmospheres, the outer boundary of the habitable zone may be extended several times as far from the star, as molecular hydrogen compensates for the loss of the incondensible greenhouse gas, via collision-induced absorption, while warding off condensation. Water and carbon dioxide may be exchanged for other greenhouse gases, which may absorb and re-radiate heat at other wavelengths or frequencies. Generally, a greenhouse gas is effective only if it is absorbent at wavelengths over which the planet is actually emitting radiation. A greenhouse gas that favors the absorption of blue light is useless if the planet emits only red light.

There is a fascinating example of alternative atmospheric chemistry in action in our Solar System. The object in question is not officially a planet, but has a fully functioning atmosphere: Titan, a moon of Saturn that is about 40 percent of the size of Earth. As in Earth's atmosphere, Titan's inert gas is molecular nitrogen. Methane is a greenhouse gas on Titan. But unlike on Earth, where methane exists only in its gaseous form, it is a condensible greenhouse gas on Titan due to the considerably lower temperatures. Instead of water, the hydrological cycle of Titan is regulated by methane. Instead of carbon dioxide, the incondensible greenhouse gas is molecular hydrogen, which plays a negligible role on Earth. Molecular hydrogen warms the atmosphere of Titan via collision-induced absorption. Titan is hardly in the habitable zone for liquid water, but it would be in the habitable zone for liquid methane! In the past, there have been several attempts to develop and launch a lander on Titan that would splash down into one of its methane lakes, but none of them have yet been approved by NASA.

Without knowledge of the major molecules of the atmosphere of an exoplanet, attempting to determine if it resides in the habitable zone for liquid water is entirely speculative. It is akin to assuming that the exoplanet has an atmosphere exactly like Earth, consisting of nitrogen, water, and carbon dioxideÑin precisely the same relative amounts, and summing up to exactly the same total mass. \textit{Declaring a freshly detected exoplanet to be in its ``habitable zone" amounts to little more than media spin if its atmospheric composition is unknown.} Even professional astronomers sometimes forget this fact.

There are further examples of how the use of the habitable zone may be an incomplete way of thinking about habitability. One of the most promising worlds to search for life in our Solar System is actually not a planet at all. Europa, one of Jupiter's moons, sits outside of the traditional habitable zone. It has no atmosphere and no surface liquid-water temperatures. A body of evidence suggests that a deep ocean exists beneath its icy surface, which may host life.  As another example, the polymath and iconoclast Thomas Gold has suggested that a hidden ecosystem exists in the crust of the Earth, which he called the \textit{deep hot biosphere}.  He hypothesized that this subsurface microbial life derived its energy from chemosynthesis rather than photosynthesis.  Gold, who was my academic great-grandfather, further suggested that the deep hot biosphere contained as much biological mass as the conventional biosphere on the surface of the Earth. Unfortunately, even if subsurface habitats for life are common on exoplanets, they would essentially be invisible to the astronomer.

\section{Thinking Outside the Zone}

Current technology largely restricts us to characterizing the atmospheres of exoplanets that are Jupiter-like in size. As technology advances, astronomers expect to be able to decipher the atmospheres of smaller, Earth-like exoplanets. Can we predict in advance, using theory, what the compositions of these atmospheres are? Unfortunately, this is a daunting task for smaller exoplanets. We expect the gas-giant exoplanets to have volatile elements (ones that vaporize at modest temperatures) in a mixture that bears some semblance to that of their parent stars. For exoplanets dominated by a rocky core, we expect the refractory elements (ones that remain solid up to fairly high temperatures) to mirror that of the star, but not the volatiles. In other words, we expect the rocks of the exoplanet to mirror the metals in the star, but not its gas.

This expectation is certainly met by Earth, whose nitrogen-dominated atmosphere hardly resembles the hydrogen-dominated Sun. On Earth, the amount of carbon dioxide present in the atmosphere is regulated by the inorganic carbon cycle, which operates on geological time scales of hundreds of thousands of years. Through the process of weathering, gaseous carbon dioxide reacts with silicate rocks and water to form calcium carbonate, which is then subducted into the Earth's mantle. This part of the cycle acts as a sink of carbon. Carbon dioxide is released back into the atmosphere via outgassing and volcanic activity. The inorganic carbon cycle acts like a geochemical thermostat: weathering is more active when the conditions are wetter and warmer, which regulates the amount of carbon dioxide present (albeit not on short enough timescales to mitigate human-induced climate change). It also means that when the Earth is trapped in a frozen, ice-covered state, weathering is shut off. Outgassing continues to increase the amount of carbon dioxide in the atmosphere of this snowball Earth, until greenhouse warming suffices to melt the ice and snow and return its climate to normalcy.

The existence of the inorganic carbon cycle on Earth suggests that to understand the atmospheres of rocky exoplanets we need to understand the geochemistry of their surfaces. Is water always the solvent? Are the minerals and rocks the same as those on Earth? Is carbon dioxide the only greenhouse gas being geochemically regulated? Are these long-term geochemical cycles necessary for stable, habitable climates? Until we resolve these puzzles, our theories will have little predictive power.

Nevertheless, Nature is offering hints that life elsewhere in the Universe, if it exists, may not be that different from what we know on Earth. Science fiction has popularized the idea of silicon-based life, as silicon resides in the same group as carbon in the periodic table of elements. This analogy breaks down when one examines the details, however. Unlike carbon dioxide, the chemical bond of silicon dioxide is too strong, while the silicon-silicon bond is too weak. Silicon dioxide (quartz) is an overly stable sink of silicon and is largely insoluble in water. These properties prevent silicon from forming a rich variety of complex molecules like carbon does. Such an expectation is consistent with what astronomers find when they point their telescopes at seemingly uninteresting parts of space: an abundance of organic molecules, ranging from methanol and glycine (an amino acid) to fullerene (the ``buckyball" with 60 carbon atoms). It appears that the building blocks of life, as we understand them on Earth, are commonly found among the cosmos---preassembled. Darwin's ``warm little pond" idea of organic molecules forming in a primeval soup on Earth may need some rethinking.

In the future, as astronomers report on the detection of molecules in the atmospheres of Earth-like exoplanets, the challenge will lie in the interpretation of the data. What are the combinations of molecules that need to be present in order for us to declare that extraterrestrial life has been detected? Classical ideas include the presence of oxygen and ozone. A potential false positive is the abiotic production of oxygen and ozone via the photolysis of water---the breakup of water molecules when exposed to ultraviolet radiation from the star. On Earth, water resides close to its surface, rendering such a process ineffective. The laws of physics, as we understand them, suggest that such a \textit{cold trap} may not be operational on all exoplanets, implying that the detection of oxygen and ozone alone cannot be robust indicators of life. (We may potentially distinguish this scenario by measuring the escape of hydrogen from the exoplanet.) It is not that oxygen and ozone are no longer useful indicators of life; it is that they cannot be the \textit{only} indicators, and we need to understand what we should additionally look for.

\section{Indicators of Life}

Another consideration in the search for life on exoplanets is that our ideas for biosignature gases are based on the atmosphere of Earth and on metabolic cycles as we understand them. It is conceivable that some fraction of rocky exoplanets may instead have thin, hydrogen-dominated, rather than nitrogen-dominated, atmospheres. In such atmospheres, atomic hydrogen acts as a radical (a reactive state where some electrons are left unpaired) and destroys most of the molecules we regard as indicators of life. Ammonia and nitrous oxide are the most promising biosignature gases in such environments, because they are spared destruction by hydrogen. Methane and hydrogen sulfide, which are produced by life on Earth, become unreliable indicators, because they may be produced abiotically via geochemistry. Clearly, whether a specific molecule can be interpreted as a biosignature gas depends on the type of atmosphere the exoplanet has.

Generally, the difficulty with making the leap from the detection of molecules in an exoplanetary atmosphere to the identification of life is that many of the gases emitted by life are also manufactured by geology. \textit{The challenge may be framed as the identification of true biosignature gases in the face of geological false positives.} Familiar gases like ammonia, carbon dioxide, methane, oxygen, and water vapor are not uniquely associated with life. Exotic ones, such as dimethylsulfide, may potentially serve as biosignature gases, but they are difficult to detect in the spectrum of an exoplanetary atmosphere because their spectral signatures are subtle.

Ultimately, the search for life elsewhere in the Universe may require that we are able to define what life actually is. There is a lesson from earlier periods of human history---for instance, when water was described by its properties, rather than by its stoichiometry, due to our ignorance (then) of chemistry. We are facing the same struggle with the definition of life, because of our current inability to frame biology in sufficiently precise terms. Astronomers will continue to improve and sharpen their search for life elsewhere, while awaiting a general definition of life---a task best left to the biologists, but one with vast cosmic implications.

\appendix

In a letter to his friend, Charles Darwin wrote, ``It is often said that all the conditions for the first production of a living organism are now present, which could ever have been present. But if (and oh what a big if) we could conceive in some warm little pond with all sorts of ammonia and phosphoric salts, light, heat, electricity, etc, present, that a protein compound was chemically formed, ready to undergo still more complex changes, at the present day such matter would be instantly devoured, or absorbed, which would not have been the case before living creatures were formed." Do the discoveries of organic molecules in space alter this picture?


\label{lastpage}

\end{document}